\documentclass[10pt,conference]{IEEEtran}
\IEEEoverridecommandlockouts
\usepackage{algorithmic}
\usepackage{graphicx}
\usepackage{textcomp}
\usepackage{xcolor}
\usepackage{xspace}
\usepackage{url}
\usepackage{multicol}
\usepackage[normalem]{ulem}
\usepackage{tabularx}
\usepackage{color}
\usepackage{colortbl}
\usepackage[switch]{lineno}
\usepackage{subcaption}
\usepackage{caption}

\definecolor{light-gray}{gray}{0.8}
\definecolor{codegreen}{rgb}{0,0.6,0}
\definecolor{codegray}{rgb}{0.5,0.5,0.5}
\definecolor{codepurple}{rgb}{0.58,0,0.82}
\definecolor{backcolour}{rgb}{0.95,0.95,0.92}

\usepackage[colorinlistoftodos,prependcaption,textsize=tiny]{todonotes}

\usepackage{enumerate}
\usepackage[shortlabels]{enumitem}

\usepackage{tcolorbox}

\usepackage{multirow}
\usepackage{pbox}
\usepackage{tabularx}

\usepackage{listings}
\usepackage{amsmath, amsthm}

\usepackage{amssymb}
\usepackage{braket}
\usepackage[ruled,vlined,linesnumbered]{algorithm2e}
\usepackage{xcolor}

\definecolor{fpbackcolor}{RGB}{242,242,242}
\definecolor{diffrem}{RGB}{202, 45, 49}
\definecolor{diffincl}{RGB}{0, 135, 90}
\definecolor{codepink}{RGB}{237, 2, 140}

\lstdefinestyle{fpstyle}{
    language = Java,
    xleftmargin = 5mm,
    xrightmargin = 3mm,
    frame=single,
    framexleftmargin=3mm,
    keywordstyle = {\color{blue}},
    keywordstyle = [2]{\color{blue}},
    keywordstyle = [3]{\color{yellow}},
    keywordstyle = [4]{\color{teal}},
    morekeywords = [3]{<<, >>},
    morekeywords = [4]{++},
    basicstyle=\ttfamily\tiny,
    commentstyle=\color{gray}\ttfamily
    alsoletter={>,?,:,!},
    morekeywords={!,>,?,:},
    morecomment=[l][\color{diffincl}]{+++},
    morecomment=[l][\color{diffrem}]{---},
    morecomment=[is][\color{diffincl}]{+*}{*+},
    morecomment=[is][\color{diffrem}]{-*}{*-},
    texcl=false
}

\newcommand{\component}[1]{\textit{#1}\xspace}

\newcommand{\parser}{\component{Parser}}
\newcommand{\alignment}{\component{Syntactic Alignment}}
\newcommand{\errorlocalizer}{\component{Error Localizer}}
\newcommand{\interpreter}{\component{Interpreter}}
\newcommand{\repair}{\component{Repair}}
\newcommand{\feedback}{\component{Feedback}}

\newcommand{\grading}{\component{Auto-Grading}}

\usepackage{cite}
\usepackage{amsmath,amssymb,amsfonts}
\usepackage{algorithmic}
\usepackage{graphicx}
\usepackage{textcomp}
\usepackage{xcolor}
\usepackage{url}
\def\BibTeX{{\rm B\kern-.05em{\sc i\kern-.025em b}\kern-.08em
    T\kern-.1667em\lower.7ex\hbox{E}\kern-.125emX}}
\begin{document}


\title{Software Engineering Educational Experience in Building an Intelligent Tutoring System
}

\makeatletter
\newcommand{\linebreakand}{%
  \end{@IEEEauthorhalign}
  \hfill\mbox{}\par
  \mbox{}\hfill\begin{@IEEEauthorhalign}
}
\makeatother

\author{\IEEEauthorblockN{Zhiyu Fan}
\IEEEauthorblockA{\textit{National University of Singapore}\\
Singapore\\
zhiyufan@comp.nus.edu.sg}
\and
\IEEEauthorblockN{Yannic Noller\IEEEauthorrefmark{2}}
\IEEEauthorblockA{\textit{Ruhr University Bochum}\\
Germany\\
yannic.noller@acm.org}
\and
\IEEEauthorblockN{Ashish Dandekar, Abhik Roychoudhury}
\IEEEauthorblockA{\textit{National University of Singapore}\\
Singapore\\
\{ashishd,abhik\}@comp.nus.edu.sg}
}


\maketitle

\begingroup\renewcommand\thefootnote{\IEEEauthorrefmark{2}}
\footnotetext{This work was done at the National University of Singapore.}
\endgroup
\thispagestyle{plain}
\pagestyle{plain}
\begin{abstract}
The growing number of students enrolling in Computer Science (CS) programmes is pushing CS educators to their limits. This poses significant challenges to computing education, particularly the teaching of introductory programming and advanced software engineering (SE) courses. 
First-year programming courses often face overwhelming enrollments, including interdisciplinary students who are not CS majors. The high teacher-to-student ratio makes it challenging to provide timely and high-quality feedback. 
Meanwhile, software engineering education comes with inherent difficulties like acquiring industry partners and the dilemma that such software projects are often under or over-specified and one-time efforts within one team or one course. 
To address these challenges, 
we designed a novel foundational SE course. This SE course envisions building a full-fledged \textit{Intelligent Tutoring System (ITS) of Programming Assignments} to provide automated, real-time feedback for novice students in programming courses over multiple years.
Each year, SE students contribute to specific short-running SE projects that improve the existing ITS implementation, while at the same time, we can deploy the ITS for usage by students for learning programming. This project setup builds
awareness among SE students about their contribution to a ``to–be–deployed" software project.
In this multi-year teaching effort, we have incrementally built an ITS that is now deployed in various programming courses. This paper discusses the \textit{Intelligent Tutoring
System} architecture, our teaching concept in the SE course, our experience with the built ITS, and our view of future computing education.
\end{abstract}


\section{Introduction}
In Computer Science (CS) education, we face the challenge of increasing student enrollments over the past few years~\cite{Singer2019}. Consequently, it has become increasingly difficult to provide high-quality and individualized learning support, particularly for novice students~\cite{mmapr2022arxiv, Mirhosseini2023}. Mirhosseini et al.~\cite{Mirhosseini2023} recently conducted an interview study with CS instructors to identify their biggest \textit{pain points}. Among other issues, they found that CS instructors struggle with limited or no Teaching Assistant (TA) support and the generally time-consuming task of providing student feedback and grading assignments. Thus, CS instructors would greatly benefit from automating tutoring activities to support TAs in their responsibilities.
Another typical problem in CS education is the provision of Software Engineering (SE) projects. Software engineering is typically a compulsory course in the university's curriculum for CS students, and it is often accompanied by development projects, in which students can collect hands-on experience in software development in a team going beyond a programming exercise. Such projects come with inherent difficulties like acquiring industry partners and the dilemma that such software projects are often under- or over-specified. Additionally, such projects are often one-time efforts within one team or one course, and students cannot experience the evolution of a software system.

In this work, we report our experience in tackling these two problems in CS education by building an \textit{Intelligent Tutoring System} (ITS) \textit{with} and \textit{for} students. As a multi-year research and teaching effort, we combine SE teaching and programming teaching via a long-term, practical, self-sustained software system that can be deployed in CS courses with programming assignments. Specifically, we use the latest research results in automated program repair (APR)~\cite{clara_pldi2018, sarfgen_pldi2018, refactory_ase2019} to generate precise patches for incorrect students' solutions. Then, we leverage the powerful natural language inference ability from large language models (LLMs) to elaborate the low-level program patch into conceptual-level feedback guidance.
  
\begin{figure}[h!]
    \centering
    \includegraphics[width=\columnwidth]{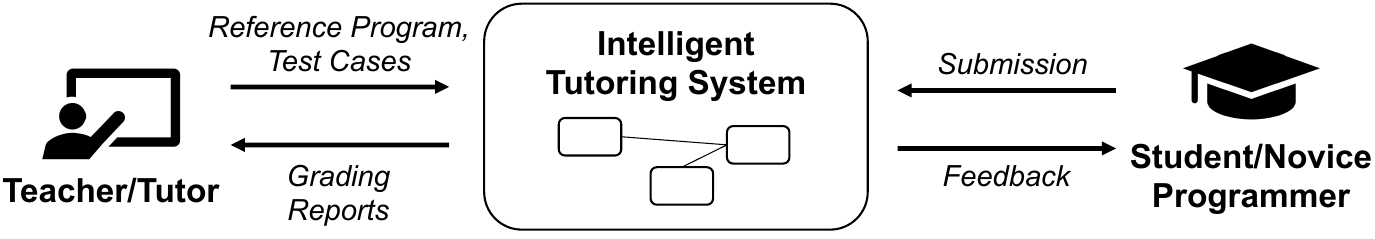}
    \caption{ General idea of an ITS that supports students and tutors in programming courses.}
    \label{fig:its-idea}
\end{figure}

Figure~\ref{fig:its-idea} shows the general idea of such a system. It can provide automated and individual feedback for student code submissions and grading support for tutors and lecturers.
Further, we involve third-year SE students in the incremental development of such a system. We offer various SE projects for the students in our advanced SE course. In this course, the students can choose from a wide range of projects, which essentially represent the development or extension of ITS components. Based on the nature of the overall project, we can conduct requirements engineering activities (e.g., surveys, interviews, and user studies) in-house because the various stakeholders are available in the university context. Each student project has the chance to contribute to the overall long-running SE project and eventually impact the learning experience of hundreds of other CS students. In our experience, this creates additional motivation because the effort is not lost, and they can relate to the users because they (at some point in their studies) also faced the challenges of learning programming.

Based on our experience with around 125 undergraduate SE students who helped develop the system throughout three years of teaching, the SE students enjoyed the course project. In particular, they liked the potential reuse of their implementation in the real deployment of the ITS. Additionally, they enjoyed the fact that there is already a system, which they have to extend (i.e., also the added complexity in understanding the already existing design and codebase).

We collected the user experience of the built ITS after each offering of our SE course. Our controlled experiments with novice students revealed that the ITS can benefit students by localizing error code snippets and highlighting the error categories. Our shadow deployment with tutors demonstrates the generalizability and quality of generated feedback, implying the potential to reduce the workload for teaching staff.

Our course not only impacts the programming courses in our university but also has the potential to impact other universities which adopt a similar teaching concept linking the teaching of software engineering with the teaching of programming. In the future, we plan to conduct more user studies to explore learning success across university boundaries.

In summary, we make the following core contributions:
\begin{itemize}[leftmargin=15pt]

\item We present our approach that facilitates both SE and programming teaching, by linking them together via developing an Intelligent Tutoring System (ITS) \textit{for} programming as a multi-semester \textit{SE project}.

\item We present our design and architecture of an extensible automated feedback system for computing education.

\item We present our long-term vision, teaching concept, and project management in the SE course that involves the incremental development of ITS.

\item We share our experience with the stakeholder's engagements through two controlled experiments with students and a large-scale shadow deployment with tutors.

\end{itemize}

\subsubsection*{Paper Structure}
We first present the research background and discuss the related work in Section~\ref{sec:background}. Section~\ref{sec:its-system} explains the overall architecture of our ITS, particularly Section~\ref{sec:feedback} and  Section~\ref{sec:autograding} highlights the system's key student-facing functionalities: feedback and grading. Then we describe our teaching concept and detailed course arrangement in Section~\ref{sec:its-teaching}.  In Section~\ref{sec:cs1-experience} and Section~\ref{sec:cs2-experience}, we report our pilot user study and experience of applying ITS in the teaching of first-year programming and second-year data structure and algorithm courses. Finally, we reflect on the challenges in organizing the course in Section~\ref{sec:challenges} and share our future vision in Section~\ref{sec:impact}.  
\section{Related Work}
\label{sec:background}

\subsection{Capstone Projects in Software Engineering Teaching}
Over the years, project-based courses~\cite{johns2013simulating,spichkova2019industry,bruegge2015software,tenhunen2023systematic,delgado2017evolving,deakin-capstone} have been applied as common sense in software engineering teaching. Students are often required to work as a team to develop software either from industrial partners or simulated real-world topics via semester-long projects.  However, there exist inherent barriers and challenges to this teaching setting. For instance, continuously collecting project topics from industry partners and establishing an efficient communication channel between stakeholders (students and company clients) are challenging tasks for the instructor. This leads to further difficulty that SE students have to either work on repetitive projects or a different, one-time effort project each year. Therefore they usually do not have a general picture of the entire system and rarely have the opportunity to experience the evolution of a mature software system.

In this work, our focus is presenting the idea of having an in-house, long-running, sustainable software engineering project in the university context.
This kind of long-running SE project shares characteristics with other community-driven course concepts \cite{source_academy_sigcse23}.
Our proposed teaching concept is novel in the sense that it links the teaching of software engineering courses and the teaching of introductory programming courses, as well as other courses like data structure and algorithms. This is done by developing an intelligent tutoring system. Students not only get training for software development but also gain exposure to the latest research in the SE community.

\subsection{Automated Feedback Generation}
\paragraph{APR-based Approach} Automated program repair (APR)~\cite{nguyen2013semfix,le2011genprog,mechtaev2016angelix} is a technique that is designed to automatically provide program patches to reduce developers' manual debugging burden. Prior research~\cite{yi2017feasibility} has shown the possibility of applying APR techniques in introductory programming courses. Over the last decade, a number of CS-1 specific APR tools have been introduced to rectify programming mistakes and provide feedback for novice programmers. AutoGrader~\cite{singh2013automated} automatically synthesizes patches for common mistakes in students' incorrect programs using manually curated program error models. Clara and SarfGen~\cite{clara_pldi2018, sarfgen_pldi2018} synthesize patches to repair students' incorrect programs at the basic-block level by editing students' faulty statements with expression ingredients from reference solutions. Refactory~\cite{refactory_ase2019} uses refactoring rules to improve repair accuracy and patch quality. Verifix~\cite{verifix_tosem2022} aims to improve the trustworthiness of generated patches by performing program equivalence verification. There are also works specifically designed for repairing syntax issues in students' submissions~\cite{TRACER,yasunaga2021break}.

\paragraph{LLM-based Approach} The emergence of LLMs has become popular in computer science education. Researchers have leverage LLMs for feedback generation in programming education~\cite{balse2023investigating,macneil2023experiences,leinonen2023using,taylor2024dcc,cambaz2024use,hellas2023exploring,kazemitabaar2024codeaid,koutcheme2024open,liu2024beyond,phung2023generating}. For example, Balse et al.~\cite{balse2023investigating} and Hellas et al.~\cite{hellas2023exploring} found that LLMs struggles to identify all issues in student's questions and false positives are common in the feedback generated by LLM. Many works~\cite {taylor2024dcc, leinonen2023using, phung2023generating} focus on generating feedback on syntax problems and error messages. While~\cite{macneil2023experiences,liu2024beyond,kazemitabaar2024codeaid, bassner2024iris} tried to build LLM-based feedback generation systems that are capable of handling general students' questions, they heavily rely on specifically curated prompts, and it remains unknown how well these systems can be adapted by worldwide educators. 

Despite these tools having shown promising results in CS-1 teaching, their research outcomes have different focuses that cannot be best utilized in a single system. In this work, we present an Intelligent Tutoring System that synergies the strengths of both APR and LLM for feedback generation of programming errors. Our modular design makes the ITS a customizable platform that can be easily adapted and evolve with the latest relevant research results over time.

\begin{figure*}[ht!]
   \centering
   \includegraphics[width=0.8\textwidth]{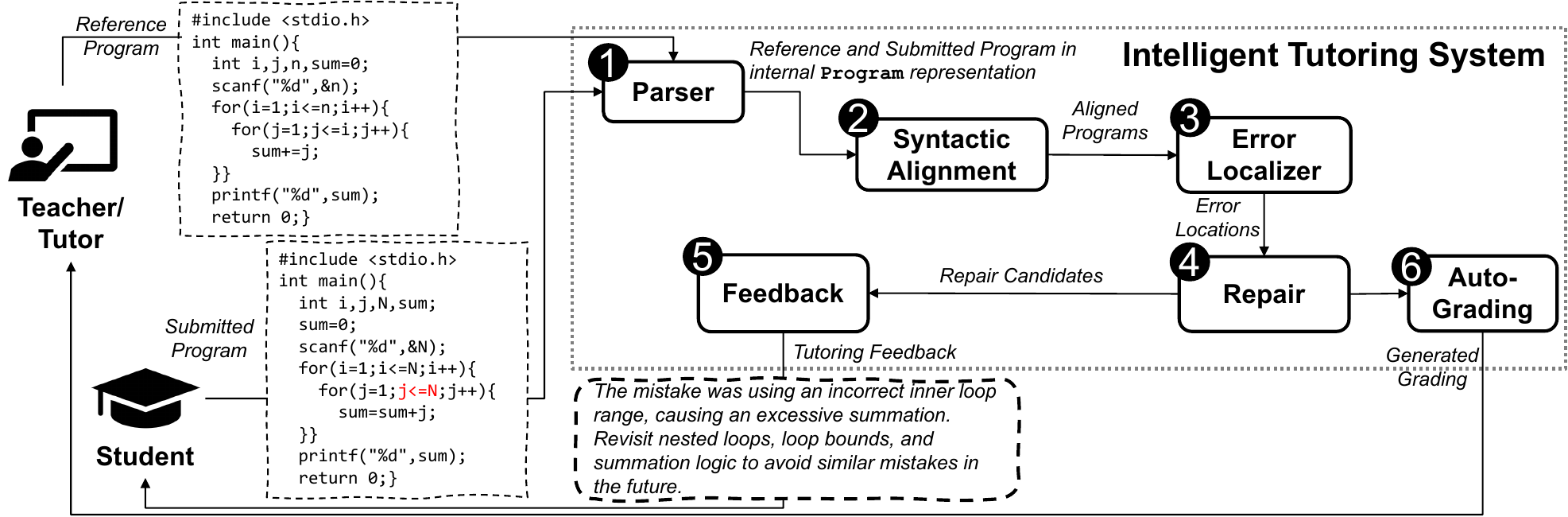}
   \caption{Illustrates the general workflow of the Intelligent Tutoring System.}
   \label{fig:workflow}
\end{figure*}

\section{Intelligent Tutoring System (ITS)}
\label{sec:its-system}

Despite prior research~\cite{yi2017feasibility, verifix_tosem2022, refactory_ase2019, clara_pldi2018, sarfgen_pldi2018, concept_maps_grading_issta23} from automated program repair and synthesis demonstrated their potential of feedback generation and grading in programming courses, these systems are not yet widely adopted in CS education. The main reasons include their prototype nature, difficulty of use, and lack of evolution. Additionally, this line of work exposes direct fixes of errors to students, which may hinder the learning process. Recent advancements in LLMs have also revolutionized automated feedback systems in computer science education~\cite{balse2023investigating,macneil2023experiences,leinonen2023using,taylor2024dcc,cambaz2024use,hellas2023exploring,kazemitabaar2024codeaid,koutcheme2024open,liu2024beyond}. These systems typically focus on prompting large language models (LLMs) for specific teaching scenarios, such as question-answering for lecture topics. However, they heavily rely on LLM output, which is known to be prone to hallucinations~\cite{balse2023investigating,hellas2023exploring}. 

In this section, we introduce the design principles of our Intelligent Tutoring System (ITS) for programming assignments that synergize the strengths of both approaches. The ITS first searches for precise bug-fixing patches with a hybrid program repair engine, and then ITS invokes LLM to use those patches as guidance to pinpoint students' conceptual misunderstandings and provide more reliable feedback. The key is to bridge the gap between accurate low-level fixing by program analysis and knowledgeable high-level explanations by LLMs. We illustrate the detailed architecture, key components, and workflow for practitioner adaptation.

\subsection{Design Principles}

To build a practical and up-to-date ITS that can be widely adopted, we adhere to the following three design principles:

\begin{itemize}[leftmargin=1em]\itemsep0em 
   \item \textbf{Language-Independent:} The ITS must be capable of processing multiple programming languages to fit the needs of various programming courses. Developing and maintaining a separate ITS for each language is both costly and impractical.
   To achieve language independence, the ITS should be designed with clean interfaces allowing language-specific plugins or adapters. 
   These plugins handle language-specific syntax and semantics, while the core system manages the general logic of tutoring and feedback. This principle ensures that the ITS can be utilized across diverse programming courses, easyily adapting to curriculum changes.

   \item \textbf{Modular and Extensible:} The ITS needs to be modular to incorporate the unique benefits of various research tools, facilitating maintenance and upgrades. The architecture should feature well-defined interfaces between modules, enabling the addition of new components or the replacement of existing ones without disrupting the overall system. This principle ensures the system’s ability to evolve by integrating the latest research findings. For instance, the core repair engine can incorporate new repair strategies, and the feedback generator can be enhanced through new interactions with large language models, such as LLM agent collaboration, making the ITS a future-oriented solution.
   
   \item \textbf{Scalable:} The ITS needs to be scalable to handle a large number of student submissions and provide feedback in a timely manner without sacrificing usability. 
   Our design of independent modules allows the dynamic deployment of scaling methods like load balancing for all components.
\end{itemize}
Inspired by prior research in program repair~\cite{clara_pldi2018, sarfgen_pldi2018, verifix_tosem2022, refactory_ase2019}, we have identified several essential components for the ITS. Figure~\ref{fig:workflow} illustrates the detailed architecture and workflow of the ITS.
The figure also includes a sample code submission with an incorrect loop condition and the corresponding generated feedback.
All components are provided via interfaces, allowing for independent implementation. In the following sections, we introduce each component in detail and explain the workflow.

\subsection{Language Parser}
\label{sec:parser}

To support multiple programming languages, we designed an internal intermediate CFG program representation that unifies different language syntaxes. This intermediate CFG program representation is capable of expressing the majority of syntax and semantics that are required in programming assignments, such as variable declarations, control structures, and basic data types. This intermediate representation ensures that the core functionalities of the ITS can operate independently of the programming language used. For example, it enables lightweight program analyses, such as control flow, variable usage, and data dependency analysis.
As the first step in the workflow, the ITS runs a grammar checker to identify the programming language of the current feedback request. Next, the parser component processes the source code of both the reference program and the student’s submission. It invokes the corresponding language-specific parser to generate the intermediate representation of the programs. This representation standardizes the code into a common format used by other components, which allows the ITS to function consistently across different languages. Currently, the parser component includes specific parsers for C, Java, and Python.

\subsection{Syntactic Alignment}
\label{sec:alignment}

One key difference between general program repair for large software and program repair for educational purposes is the availability of an expected program specification in the form of a reference implementation. The \alignment component is designed to align the reference program with the student’s submission. It processes the intermediate representations of both the reference and student programs to identify matching basic blocks and map the existing variables for each function within the programs. The alignment algorithms~\cite{clara_pldi2018, refactory_ase2019, verifix_tosem2022} are based on the similarity of control flow and variable usage, specifically using Def-Use Analysis, to compare the reference and student programs. The results of this alignment can then be used to pinpoint the locations where the reference and submitted programs diverge in behavior. Furthermore, this information is instrumental in attempting to repair the submitted program by leveraging the data from the reference solution. Note that, the ITS can take in multiple reference solutions with different solving strategies to increase the alignment success rate like existing APR tools~\cite{verifix_tosem2022,refactory_ase2019}.

\subsection{Error Localizer and Interpreter}

\label{sec:errorlocalizer}

Error localization is a crucial step in APR systems that aims to identify the buggy locations within the software. In the context of programming education, error localization identifies specific basic blocks or expressions that violate the expected specifications. The \errorlocalizer component employs several dynamic execution-enabled localization algorithms to trigger erroneous behavior in the student’s program. These algorithms include trace-based localization and statistical fault localization~\cite{wong2016survey}. The dynamic program execution is facilitated by an \interpreter component. This interpreter allows the execution of a program in its intermediate CFG-based representation without the need for compilation or execution on the actual system. It generates an execution trace with the sequence of executed basic blocks and a memory object, which holds the variable values at specific locations. The \errorlocalizer component utilizes the \interpreter to execute test cases while observing the variable values at specific locations. This process enables the system to detect semantic differences between the reference and submitted programs, thereby pinpointing the precise locations of errors.

\subsection{Repair Engines}
\label{sec:repair}

Given the reference programs, student submissions, and the identified error locations as input, the \repair component attempts to fix the submitted programs by generating edits that transform the student’s program to be semantically equivalent to the reference program. The \repair component acts as an engine that can utilize various repair strategies, such as optimization-based repair~\cite{clara_pldi2018}, synthesis-based repair~\cite{refactory_ase2019, verifix_tosem2022}, and LLM-based repair~\cite{mmapr2022arxiv}. Upon receiving a repair request from the previous components, these repair strategies are invoked in parallel to search potential repair candidates for all identified errors (even with multiple errors existing in a student submission). Then, the repair engine selects the optimal repair candidate that minimally alters the student's submission. This approach aims to rectify students' mistakes while preserving their original intentions as much as possible. Note that the repair candidate is managed at the level of the intermediate representation of the program, and we convert it back to the source code before proceeding to the feedback generation. 

\subsection{Feedback Generator}
\label{sec:feedback}

With the collected information from previous components, the \feedback component generates natural language explanations to guide students in correcting their mistakes without revealing the direct answer. This component incorporates a common front-end prompt interface with various LLM backends, allowing flexible switching between different LLMs and easy integration of new LLMs. Currently, it supports both commercial LLMs like GPT and Claude series, as well as open-source LLMs like LLaMA~\cite{touvron2023llama} from Meta. We use GPT-3.5-turbo as the default LLM backend to balance performance and cost.
Our example prompt template consists of (1) a task description, (2) the student submission, and (3) program patches from the repair engine annotated with error locations:
\begin{tcolorbox}[boxrule=1pt,left=1pt,right=1pt,top=1pt,bottom=1pt]
\footnotesize
\noindent You are a teaching assistant for an introductory programming course. \\

You will be given 
(1) text description of a programming task
(2) a wrong student submission
(3) sample fixes to the wrong submission.
\\

Based on the sample fixes, please explain to the student conceptually why the mistake exists in this task, and what programming concepts the students should revisit.
\\

Description of the programming task: \{description\}
\\

Wrong student submission: \{student code\}
\\

Fixes to the wrong submission:
\{patches from repair engine\}

\end{tcolorbox}

These ingredients in the prompt can be seen as a precise hint that instructs the LLM to generate feedback that highlights both assignment-specific mistakes and related general programming concepts. This dual focus helps students understand the underlying issues more comprehensively. Note that, this feedback generator is able to integrate any latest prompt engineering methods in computing education and is customizable for specific assignment categories.

\subsection{AutoGrader}

\label{sec:autograding} 
Test-suite based automated grading suffers from the problem that a small mistake by the student can cause many test cases to fail. To provide better support for tutors, we integrate recent research in \textit{conceptual} auto-grading into the ITS. This technique aims to assess the conceptual understanding of the student and awards grades accordingly~\cite{concept_maps_grading_issta23}.
This is achieved by constructing a concept graph from the student's attempt and comparing it with the concept graph of the instructor's reference solution. It automatically determines which of the ingredient concepts being tested by the programming assignment are correctly understood by the student. Given the instructor-provided reference solutions and students' incorrect solutions, 
the \grading component generates a grading report for the tutor. It assesses the student's submission by their missing or improperly used programming concepts to address the over-penalty issue~\cite{concept_maps_grading_issta23} in the conventional test-based assessment.

Overall, the designed ITS serves as a platform that can continuously integrate the latest research efforts from APR and LLM for CS education.
\section{Design of Software Engineering Course}
\label{sec:its-teaching}

This section presents the design of our new variant of SE courses. This course simultaneously enhances SE and programming teaching by embedding the \textit{Intelligent Tutoring System} discussed in Section~\ref{sec:its-system} as a project within the curriculum. We outline the high-level teaching concept, provide an overview of the long-running ITS SE project, and detail the specific course structure and short-running project setup.

\subsection{Teaching Concept }
\label{sec:concept}

Traditional software engineering (SE) courses teach students basic SE practices by creating similar small-scale, one-time effort projects. Instead, our teaching concept highlights the ``\textit{foundations}'' in two dimensions: the foundational research in SE and the foundational principles of software development. We aim to provide senior SE students with (1) exposure to frontier research ideas, such as fuzzing, debugging, static analysis, and program repair, and (2) an immersive software development environment of contributing to an existing, functional, in-use codebase that allows the students to go beyond \textit{programming-in-the-small} in the course project. 

Our SE course achieves this vision by linking software engineering and programming education together. We embed real-world \textit{Intelligent Tutoring System} development as an SE project within the course curriculum.
This project is well-framed and particularly interesting in the university context because (1) the demand for programming tutoring support will exist for a long time, and we can continuously collect feedback from users to curate new requirements. The user feedback can serve as project topics for the next iteration of our SE course. (2) The SE students who contribute to ITS can relate to the end users since they once had to learn programming, (3) all the related stakeholders are available in the university, which enables requirement elicitation and milestone discussions, (4) the techniques required for ITS is highly relevant to advanced SE skills. For instance, program analysis, software design, program verification, and program repair.

\begin{table*}[ht!]
\caption{Course assignments that accompany the major project milestones.}
\label{tab:assignments}
\footnotesize
\centering
\begin{tabular}{l|l|l}\hline
\textbf{ID} & \textbf{Topic} & \textbf{Details}  \\
\hline  
1 &
Requirements Analysis Elicitation &
\pbox{9cm}{Preparations and questions for the interview session with the customer.}
\\ \hline

2 &
Requirements Modeling &
\pbox{9cm}{Requirement modeling with UML Use Case and Activity diagrams.}
\\ \hline

3 &
{Architectural Drivers and Variants} &
{Discussion of architecture variants and the requirements that influence architectural design.}
\\ \hline

4 &
Strategy and Project Planning &
\pbox{9cm}{Project-specific planning including a Gantt-Chart and a resource plan.}
\\ \hline

5 &
\pbox{3cm}{Detailed Design }&
Structural and behavioral design of the students' implementation with UML models
\\ \hline

6 &
Intermediate Deliverable &
\pbox{9cm}{Towards the middle of the course, we ask the students to submit a minimal project implementation and a report with their project plans and various models.}
\\ \hline

7 &
Validation (i.e., Unit Testing) &
\pbox{9cm}{Test case design and test report.}
\\ \hline

8 &
Presentation \& Final Artifact&
At the end of the course, all teams need to present their project and submit their code.
\\ \hline

9 &
\pbox{3cm}{Final Report} &
\pbox{9cm}{After the presentation, the students additionally need to submit a final report, including a retrospective of their project and design decisions.}
\\ \hline
\end{tabular}
\end{table*}

\subsection{Overview of Long-running ITS Project}

Figure~\ref{its-fig:concept} illustrates the overall diagram of the Intelligent Tutoring System as a self-sustained long-running SE project that evolved over multiple years in our SE course. 

\begin{figure}[h!]
    \centering
    \includegraphics[width=0.9\columnwidth]{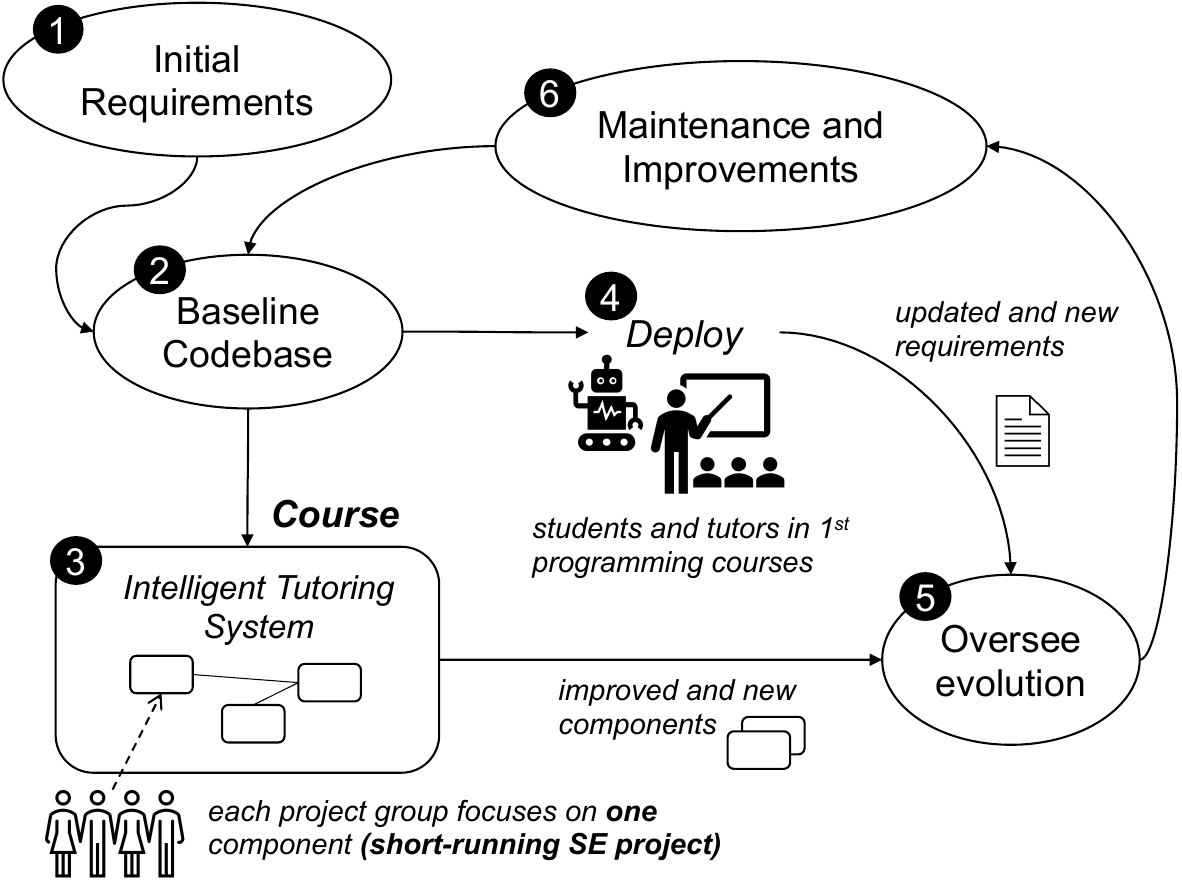}
    \caption{Concept of the long-running ITS project that is incrementally built and improved by short-running projects inside SE teaching environment.}
    \label{its-fig:concept}
\end{figure}
\setlength{\textfloatsep}{7pt}

\subsubsection{Initial Requirement Collections} 
Before we started any development activities on ITS, we collected initial requirements based on the prior research experience of automated program repair~\cite{sarfgen_pldi2018,clara_pldi2018,singh2013automated, concept_maps_grading_issta23,refactory_ase2019} in educational scenarios and discussions with first-year programming course lecturers. At a high level, the initial requirements are tutor-oriented, which mainly consists of highlighting suspicious error code snippets, producing precise code patches, and drafting high-level feedback explanations on behalf of tutors. 

\subsubsection{Codebase and Architecture Design} In step 2, we developed a baseline codebase, which included designing the artifact and the desired workflow discussed in detail in Section~\ref{sec:its-system}. This first version already defined interfaces between components and provided common data structures. The baseline also included a prototypical implementation for most of the initially planned components to test their feasibility. 
Having an initial baseline codebase provides the students with additional requirements like the existing architecture, which should not be changed. It also provides them with an environment to work on a project with partial existing functionalities.

\subsubsection{Course Projects Setup} For the SE course project (step 3), we design multiple short-running SE projects based on the feedback from first year course instructor in the requirement elicitation session of our course, and these projects essentially represent the implementation variants of existing or new components. 
We discuss those short-running projects in detail in Section~\ref{sec:short-running}.

\subsubsection{Deployment and Refinement} After each SE course, our teaching team evaluates all projects and integrates the best contributions of each project topic into our baseline implementation (step 5). Therefore, over the years, the baseline will grow and improve. At the same time, we also deploy the increments of the system in real-world programming courses 
and collect additional feedback and requirements from students and tutors (step 4). To keep the implementation standards high and to ensure that our architecture and design can cope with the increasing codebase and the possibly new and changing requirements, we constantly maintain and improve the implementation (step 6).

Overall, our  Intelligent Tutoring System as a long-running SE project course is structured so that the teaching of SE projects is accomplished over multiple years via a real-life SE project. Over the years, the ITS became more and more robust, and varied, with the continuous effort of each year's students' contributions. It eventually became a full-fledged functioning automated tutoring system that is being deployed.

\begin{table*}[h!]
    \centering
    \tabcolsep5pt 
    \caption{Examples of provided Short-running projects and trained skills in our SE course over the years.}
    \label{tab:req-evo}
       \begin{tabular}{l|l}
        \hline
        \textbf{Project} & \textbf{Trained Skills}\\\hline
        Program Parser & Develop an understanding of AST and CFG. Convert program from source language to intermediate representation.
        \\\hline
        Program Alignment & \pbox{11cm}{Develop skills in static analysis (e.g., Def-Use) and practice them in aligning two programs.}
        \\\hline
        Interpreter & Develop a program interpreter that supports ITS's intermediate representation. Experience dynamic program analysis.
        \\\hline
         Error Localization & {Understand and implement frontier error localization research (e.,g execution-based, statistical-based fault localization.)}
        \\\hline
        Repairing Engines & \pbox{11cm}{Understand and implement frontier various APR research~\cite{verifix_tosem2022,clara_pldi2018,sarfgen_pldi2018,refactory_ase2019}, the approaches include program verification, synthesis, ILP optimization, and LLM-assisted repair~\cite{mmapr2022arxiv}.}
        \\\hline
        Feedback Generation & \pbox{11cm}{Explore LLM application in explaining low-level program patch to conceptual-level guidance.}
        \\\hline
        \pbox{2cm}{Tests Generation} & Using fuzzing or mutation testing to generate incorrect programs that test the capability of the whole ITS.
        \\\hline
    \end{tabular}
    \label{tab:projects}
\end{table*}

\subsection{Overview of SE Course Management}
\label{sec:se-course}

\subsubsection{Course Curriculum}
\label{sec:curriculum}
The course curriculum focuses on the main activities in SE.
Furthermore, we introduce students to selected relevant SE topics for our project, e.g., automated program repair, static analysis, and fault localization. 
Each lecture is separated into two parts: (a) the teaching of foundations in the aforementioned areas, and (b) the teaching of project-specific knowledge and corresponding applications.

\subsubsection{Requirements Analysis and Modeling}
The course starts with a focus on requirements engineering, their elicitation, and modeling. Therefore, we invite stakeholders like lecturers and tutors from the first-year programming courses to an interview session with the third-year students. This interview session is prepared with corresponding assignments about question design and followed up with requirements modeling exercises using UML Use Cases. We also teach other requirements modeling, e.g., with finite state machines and sequence diagrams.

\subsubsection{Software Architecture and Design}
Afterwards, we introduce general principles for software architecture design and modeling. The project-specific part of the lecture introduces the existing architecture and its components, including the available interfaces, which need to be used by the students in their own implementations. We further discuss architecture variants of the existing architecture to discuss pro and contra of the made design decisions.
Our baseline Java implementation already provides the students with elementary classes and functionalities, which they can and need to reuse. To illustrate the fine-grained design, we first introduce relevant design principles and patterns that occur in our implementation. We do not give a comprehensive introduction to design patterns because there is another dedicated software design course in our institution. Instead, we only introduce the most relevant design aspects to enable the students to work on the projects.

\subsubsection{Project Planning and Implementation}
As part of the assignments, the students have to submit a project plan. Therefore, we also introduce the basics of project planning, work package design, and milestone and resource planning, including necessary models like Gantt-Charts. The coding itself is a major part of the project and is mostly supported by the mentors in project-specific meetings. The lecture introduces general principles like Clean Code and testing and debugging techniques meant to help the students in their concrete implementation efforts.

\subsubsection{Testing, Debugging, and Integration}
As automated testing and debugging is a major part of an intelligent tutoring system, we also introduce several validation concepts and debugging techniques. In particular, we teach foundations in test-suite estimation, functional testing, whitebox testing, structural testing, dataflow testing, and mutation testing. To this end, we also introduce the basics of static analysis like control-flow graphs (CFGs) and Define-Use Analysis (DUA). Furthermore, we discuss the basics of debugging with the TRAFFIC principle and delta debugging and dive deeper into the basics of static and dynamic slicing and statistical fault localization. Towards the end of the curriculum, we also discuss integration testing strategies and the related challenges.

\subsubsection{Project-Specific Topics}
In addition to the foundations in general software engineering, we teach the background in automated program repair and provide an overview of existing solutions for ITS components. Depending on the advertised projects, we also discuss more specialized topics like taint analysis and Worst-Case Execution Time (WCET) analysis to ensure the students have the relevant background and material to work on their projects.

\subsubsection{Labs and Assignments}
Each week in our curriculum is accompanied by a lecture and a lab session. The labs are used to meet in smaller groups of students and discuss their assignments. The assignments track the major milestones in the student's projects. We share the
course assignment overview in Table~\ref{tab:assignments} for practitioners.

\subsection{Overview of Short-running Projects}
\label{sec:short-running}

\subsubsection{Project Preparation} As the key part of the SE course, we carefully curated a set of short-running projects before each semester started. Those short-running projects are not only engineering efforts but also cover different research topics in the Software Engineering community. Table~\ref{tab:projects} shows a few examples of short-running projects that were provided over the years. The topics range from program structure understanding, static analysis practice, replication of error localization, and automated program repair techniques. In addition, we prepared a specific testing project that helped students gain fuzzing and mutation testing experience.
In the first year, we mainly had projects to build program analysis capabilities. We further designed projects to extend core features like \textit{Automated Feedback}, \textit{Automated Grading}, and \textit{Automated Repair} in the second year. Later, we also encouraged students to integrate LLMs, so the ITS can synergy existing program analysis artifacts with generative AI.
These projects are inherently different from traditional SE courses that merely focus on development activities. The additional context on SE research exposed students to the fundamental techniques behind software artifacts.

\subsubsection{Team Management and Project Guidance}

To reduce students' workload, we ask the students to form groups of 3-4 people to work on the project. We allow them to search for their team members instead of a random assignment by the teaching team. We prepare an ungraded \textit{Assignment 0} for the project selection, which provides an overview and additional references for all available projects for the specific year. Each team can bid for three projects, while the teaching team allocates the final project. This is to avoid most of the students working on the same topics.
Additionally, we assigned each team a graduate-level mentor who are familiar with the project topics to help students get started on their project smoothly. Each team was required to meet and discuss weekly with their mentor focusing on the team's planning, design, and implementation progress. Interestingly, those graduate mentors all had experience in tutoring programming courses. So the deep involvement of graduate mentors also works like discussion meetings with stakeholders.

Through the course and project organization, the students advance their skills in software development,  grasp a deeper understanding of fundamental SE concepts, and expose them to frontier SE research. All student projects eventually contribute to an ITS, whose details were discussed in Section~\ref{sec:its-system}.

\section{Experience in CS-1 Teaching}
\label{sec:cs1-experience}

We evaluated the Intelligent Tutoring System (ITS) after two semesters of development in a first-year programming course through two methods: (1) a pilot study with 15 novice students from the course, and (2) a shadow deployment where we compared ITS-generated feedback to that of human tutors.

\subsection{Pilot Study}

\subsubsection*{Study Methodology} 

\begin{figure}[h!]
\centering
\includegraphics[width=0.8\linewidth]{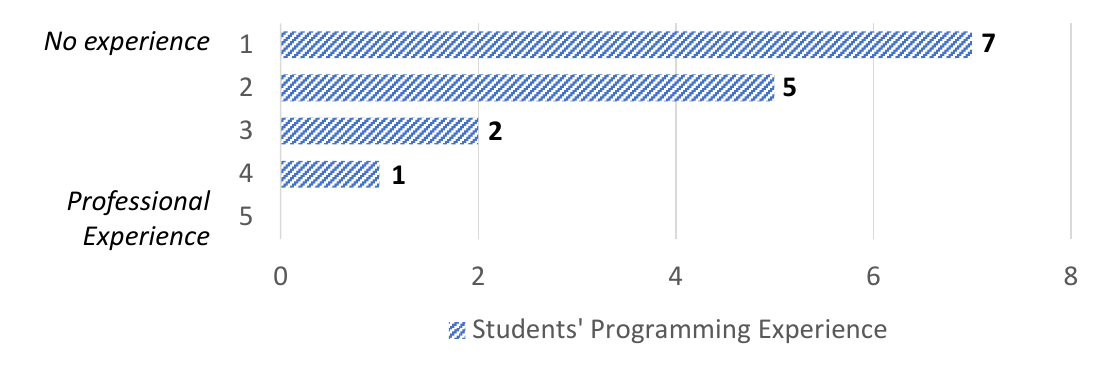}
\caption{Participants' Self-Assessed Experience}
\label{fig:background}
\end{figure}

In the pilot study, we randomly divided participants into two groups, Group A and Group B, based on their self-assessed programming experience (shown in Figure~\ref{fig:background}). Group A had access to the ITS, while Group B did not. All participants were instructed to complete four programming tasks using our Learning Management System (LMS), which allowed them to run provided test cases and submit solutions multiple times within a 20-minute time limit per task. The study was structured into three parts: (1) a background survey, (2) the programming tasks, and (3) a feedback survey. After completing their tasks, Group B was introduced to the ITS to offer their feedback, allowing us to collect input from both groups on the system’s potential impact on learning experiences.

\subsubsection*{Programming Tasks}
We selected four entry-level programming tasks from past mid-term exams of the CS-1 course, each covering a fundamental programming concept. The tasks included: (1) removing duplicates from a tuple, focusing on loop and tuple manipulation; (2) iteratively reversing a string, covering loop and string manipulation; (3) recursively reversing a string, introducing recursion; and (4) iteratively reversing an integer, practicing while-loop usage. These tasks were chosen to represent practical challenges that students commonly encounter in early programming courses.

\subsubsection*{Result Analysis for Students}
\label{sec:stu-results}
We recorded the submitted solutions and timestamps for each non-duplicate attempt on the programming tasks. A task was considered \emph{solved} if a student’s solution passed all test cases. Across both groups, we received 128 attempts in total --- 65 from Group A and 63 from Group B.
For open-ended questions, we applied qualitative content analysis using coding methods based on Schreier’s approach \cite{schreier2012qualitative}. One author conducted the initial analysis, and another reviewed the codes. After discussion, we finalized the analysis.

\begin{table}[!h]
\setlength{\tabcolsep}{2.8pt}
\setlength{\extrarowheight}{1pt}
\caption{The average number of failed attempts, rectification rates, average rectifying time of failed attempts in minutes.
}
\label{tab:stu-quantitative}
\footnotesize
\centering
    \begin{tabular}{l|cc|cc|cc}\hline
        \textbf{Tasks} & \multicolumn{2}{c|}{\pbox{1.5cm}{\textbf{Avg \# Failed Attempts}}} & \multicolumn{2}{c|}{\pbox{1.5cm}{\textbf{Rectification Rate}}} & \multicolumn{2}{c}{\pbox{1.5cm}{\textbf{Avg Rect. Time (mins)}}}  \\
         & A & B & A & B & A & B  \\
\hline
Task 1 & 4.8 & 4 & 4/5 & 0/2 & 7 & - \\\hline
Task 2 & 1.9 & 5.5 & 7/7 & 3/4 & 9.2 & 9.3 \\\hline
Task 3 & 2.3 & 2.8 & 5/5 & 2/4 & 4.6 & 2.5 \\\hline
Task 4 & 2.3 & 3.1 & 5/6 & 5/7& 4.5 & 11.3 \\\hline
Total & 2.7 & 3.7 & 21/23 & 10/17 & 6.7 & 8.9  \\\hline

\end{tabular}
\end{table}

\subsubsection*{Fewer attempts, higher accuracy}
We then analyzed students’ performance, focusing on failed attempts. Table~\ref{tab:stu-quantitative} provides an overview of the average failed attempts, rectification rates, and time taken to correct solutions across both groups. The rectification rate (X/Y) refers to the number of students (X) who eventually corrected their solutions, out of the total number of students (Y) who initially failed on a task.
As shown in Table~\ref{tab:stu-quantitative}, students who received assistance from the ITS (Group A) solved more tasks with fewer overall attempts compared to those without ITS (Group B). Although Group A made slightly more attempts on Task 1, it’s important to note that none of the students in Group B who failed this task were able to rectify their solutions. This suggests that Group B’s fewer average attempts may be due to a lack of understanding of how to correct their mistakes, leading them to give up after a few tries.
On average, Group A made 2.7 failed attempts, compared to Group B’s 3.7. While this difference is small, Group A showed a much higher success rate in correcting their errors, successfully resolving 21 out of 23 failed attempts (91.3\%). In contrast, Group B struggled more after failing their first attempt, managing to fix only 10 out of 17 failed attempts (58.8\%). This highlights the effectiveness of the ITS in guiding students through the correction process.

Regarding rectifying time, Group A was also faster, with an average of 6.7 minutes to fix one incorrect solution, compared to Group B's average of 8.9 minutes. The average rectifying time for task 1 in Group B is unavailable as none of the students could rectify their incorrect attempts. Moreover, the average rectifying time for Group B is significantly lower for Task 3 (2.5 minutes) because the two incorrect solutions were almost correct (e.g., typos).

\subsubsection*{Usefulness of ITS}

\begin{figure}[h]
\centering
\includegraphics[width=0.9\linewidth]{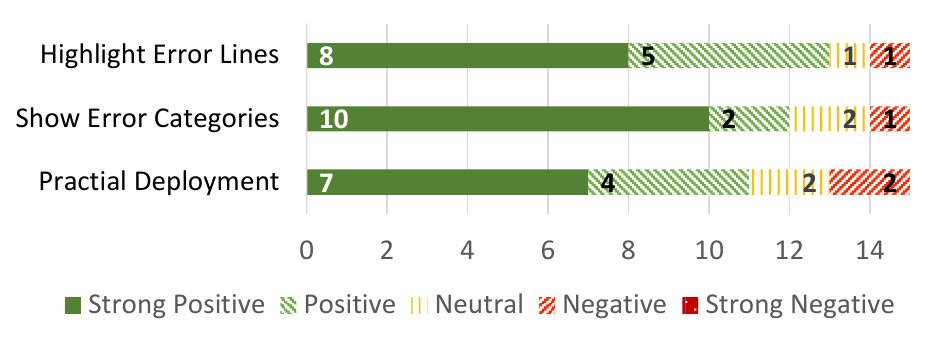}
\caption{Students' feedback of ITS}
\label{fig:stu-result}
\end{figure} 

Figure~\ref{fig:stu-result} presents the feedback from students on the usefulness of the ITS, focusing on features such as highlighting potential error lines in the code editor and providing hints about error categories.
The survey results indicate that most Group A students found the ITS helpful and were satisfied with its feedback and functionality. Over 80\% of the students responded positively to the highlighted lines and mistake categories, finding them useful for improving their code. Additionally, more than 73\% would like to have the ITS deployed in their programming courses.

\subsection{Shadow Deployment with Tutors}
\label{sec:deployment}

In addition to the controlled experiments, we ran a shadow deployment of ITS in a first-year programming course with 571 students during the fall semester of the 2023/2024 academic year. The goal is to compare ITS-generated feedback with the tutor-curated feedback before deliver to students.
In this first-year programming course, students submit their solutions on an LMS that automatically runs pre-defined test cases for programming assignments. 
After the deadline, tutors manually review incorrect submissions and write personalized feedback to students who submit incorrect solutions.

We deployed ITS for 30 programming tasks spanning six weeks of assignments to automatically generate feedback for submitted solutions that failed test cases. Note that, we excluded the first two introductory weeks and the final week, which involves Object-Oriented Programming (OOP) that the ITS did not support yet. Additionally, we did not generate feedback for empty solutions.

Throughout the semester, we collected non-empty 1,835 incorrect solutions. The deployed ITS successfully generated semantically correct patches for 1,758 (95.8\%) incorrect submissions by its repair engine. A patch is deemed correct if it makes the original submission pass all test cases for a particular programming assignment. Then the feedback component generated corresponding natural language comments for these submissions. 
Furthermore, we randomly sampled 10\% of ITS-generated feedback to manually evaluate the quality of ITS's automated feedback by assessing whether they were semantically equivalent to the corresponding tutors' feedback. Our manual analysis shows that 136 (77.2\%) of ITS-produced feedback is semantically equivalent to tutors'  feedback, illustrating its capabilities to assist human tutors.

\section{Experience in Data Structures Teaching}
\label{sec:cs2-experience}

In parallel to the shadow deployment with tutors, we explored the feasibility of enhancing ITS to handle more advanced tasks such as Data Structure and Algorithms in our SE course. Specifically, we provided SE projects to add support for OOP and advanced LLM-based repair engines~\cite{mmapr2022arxiv}, which were initially missing. These additional functionalities enabled the ITS to define and operate on data structures and improved its ability to fix complex errors. We share our experiences through an anonymous controlled experiment with 30 second-year students from a data structure and algorithm course to demonstrate the evolving status of the ITS project.

This controlled experiment consists of four LeetCode tasks --- two on Tree and two on Graph topics. These tasks (shown in Figure~\ref{fig:user-study}) have similar difficulties and they are closely relevant to the students' weekly problem sets. We conducted the controlled experiment at the end of the semester, and 
\begin{figure}[h!]
\begin{subfigure}[h]{\columnwidth}
\begin{tcolorbox}[colback=white, colframe=gray!90, boxrule=1pt,left=1pt,right=1pt,top=1pt,bottom=1pt, title=Tree Topic, fontupper=\footnotesize, fonttitle=\footnotesize]
 We assess students’ understanding of \textbf{Binary Search Tree definition and Pre/In/Post Tree Traversal}, which is covered in Week 4.\\
 Task 1: "Construct Binary Tree from Preorder and Inorder Traversal"\\
Task 2: "Construct Binary Tree from Inorder and Postorder Traversal"
\end{tcolorbox}
\vspace{1mm}
\end{subfigure}
\begin{subfigure}[h]{\columnwidth}
\begin{tcolorbox}[colback=white, colframe=gray!90, boxrule=1pt,left=1pt,right=1pt,top=1pt,bottom=1pt, title=Graph Topic,
fontupper=\footnotesize, fonttitle=\footnotesize]
 We assess students’ understanding of the \textbf{Shortest-Path Problem in Graph}, which is covered in Week 8.\\
 Task 3: "Minimum Cost to Reach City With Discounts"\\
Task 4: "Minimum Cost to Buy Apples"
\end{tcolorbox}
\end{subfigure}
\caption{Overview of the Tasks Used in Control Experiment.}
\label{fig:user-study}
\end{figure}
we equally divided the 30 student participants into two groups based on their performance in weekly assignments.
Group A has access to ITS feedback (after finalizing submissions) for Tasks 1 and 3. Therefore, we explore whether this feedback helps them to solve Tasks 2 and 4 (which are similar to Tasks 1 and 3, respectively). For example, additional feedback could strengthen their conceptual understanding of the topic. Group B has no access to ITS feedback for any of the four tasks.
Each task had a time limit of 25 minutes.

Figure~\ref{fig:2040-result} shows the number of correct submissions for each task. For Task 1 and Task 3, both Group A and Group B performed similarly, with a 30\% correct submission rate. However, for Task 2 and Task 4, Group A’s correct submission rate increased to 60\%, compared to 40\% for Group B. Although some of this improvement may be attributed to task similarities and additional given time, the increase in correct submissions for Group A suggests that the ITS feedback could have had a positive impact on their understanding. While these differences are not statistically significant, they indicate a promising trend in favor of ITS’s potential to support students’ learning.

 \begin{figure}[h!]
    \centering
    \includegraphics[width=\linewidth]{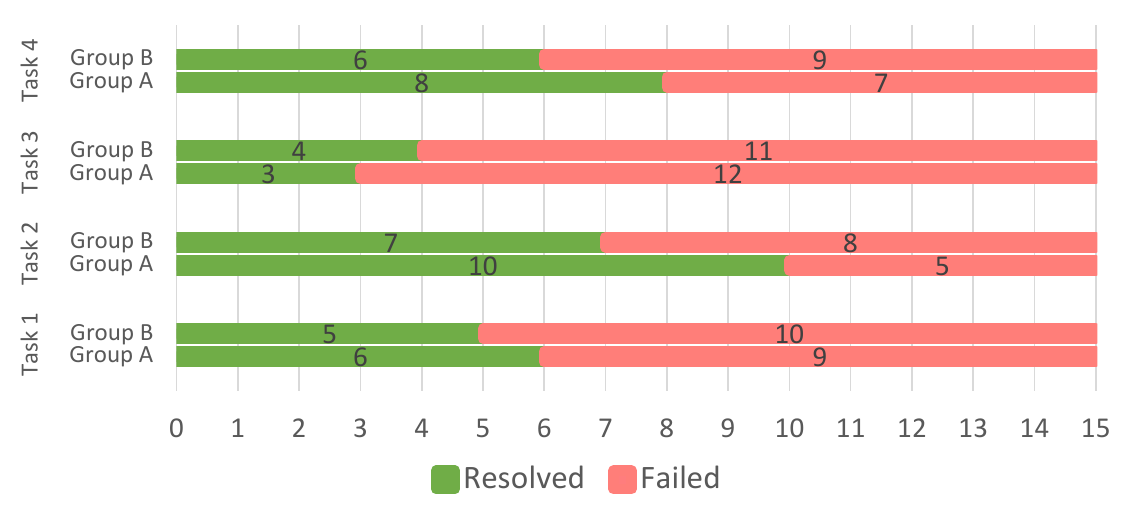}
    \caption{Number of Correct Submissions on the Four Tasks.}
    \label{fig:2040-result}
\end{figure}

We present a case study of a Group B student who was close to solving Task 1 and Task 2 but ultimately failed both. The student’s Task 1 submission is shown in Figure~\ref{fig:demo-incorrect-2}. At first glance, the code appears correct; however, the issue lies in the “build” helper function at lines 14–15. The redundant condition (if (high == low)) causes the code to skip recursion for the root node, resulting in incorrect tree construction.

The student made 8 attempts on Task 1, arriving at a nearly correct solution that only required a small fix—removing the redundant condition. Given that Task 2 is isomorphic to Task 1, we expected the student to solve it within the additional 25 minutes provided. However, the student reused their Task 1 solution, submitting 13 attempts without modifying the faulty condition.
This example demonstrates how even students who are close to the correct solution can struggle without timely guidance. We show the ITS-generated feedback (Figure~\ref{fig:demo-feedback}) to this student after the experiment, the student immediately identified and corrected the error in both tasks. This highlights the importance of immediate feedback in supporting students when they are near the correct solution but lack the insight to make the final adjustment.

\begin{figure}
   
\begin{subfigure}[h]{\columnwidth}
\begin{lstlisting}[language=Java, basicstyle=\tiny, style=fpstyle]
public TreeNode buildTree(int[] preorder, int[] inorder) {
    if (preorder.length <= 0 || inorder.length <= 0)
        return null;
    if (preorder.length != inorder.length)
        return null;
    return build(preorder, inorder, 0, inorder.length - 1);
}

public int index = 0;
private TreeNode build(int[] preorder, int[] inorder, int low, int high) {
    if (high < low) 
        return null;
    TreeNode root = new TreeNode(preorder[index]);
--- if (high == low)
---     return root;

    int mid = search(low, high, inorder, preorder);
    
    index++;
    root.left = build(preorder, inorder, low, mid - 1);
    root.right = build(preorder, inorder, mid + 1, high);

    return root;
}
\end{lstlisting}
\caption{Example Incorrect Submission from Group B for Task 1.}
\label{fig:demo-incorrect-2}
\vspace{5mm}
\end{subfigure}

\begin{subfigure}[h]{\columnwidth}
\begin{tcolorbox}[colback=white, colframe=gray!90, boxrule=1pt,left=1pt,right=1pt,top=1pt,bottom=1pt, title=Feedback from ITS,fontupper=\footnotesize, fonttitle=\footnotesize]
Mistake on Termination Condition: This submission prematurely skips the recursive calls and returns the root node before the subtrees are properly constructed. By returning prematurely, the build function does not explore the entire preorder or inorder arrays, leading to an incomplete tree construction.
\end{tcolorbox}
\caption{Feedback for Figure~\ref{fig:demo-incorrect-2} Curated by ITS.}
\label{fig:demo-feedback}
\end{subfigure}
\caption{Example of an Incorrect Submission from a Group B Student in Task 1 and ITS-generated feedback.}
\label{fig:demo-ds}
\vspace{5mm}
\end{figure}
\section{Challenges \& Lessons Learned}
\label{sec:challenges}
To further share our experience with our combined research and teaching effort, we report the challenges we faced and the lessons learned concerning the teaching of our advanced software engineering course.



\subsection{Incentives for Stakeholders}
We have three main user groups: the students who receive feedback, the tutors who can use the ITS to better understand the students' errors and get grading support, and the lecturers who provide the inputs like assignments and reference implementations.
\textit{Lecturers} are naturally concerned about deploying more tools, including the potential negative effects on the learning outcome caused by inaccurate output. To gradually convince the lecturers, we decided to first focus on a targeted shadow deployment for tutors. For tutors, an imperfect output is less critical and still can provide helpful guidance to them and help us to get feedback continuously.
To engage with \textit{first-year students}, we designed a user study that not only has a monetary reimbursement but also provides additional programming training and an extra tutorial after the user study to explain the programming tasks to them individually.
%
The \textit{third-year students} who develop the components in our course showed great interest in our project because it is (or will be) deployed in a real context and because they like working on a larger project with existing parts. Overall, it is a valuable experience for them, as shown by the following student quotes about the question of what they liked the most in the course:

\begin{center}
    \textit{``As the module is new, its content to be taught may change but I'm certain the ITS project is here to stay."}
\end{center}
\begin{center}
    \textit{``The project component – It's really interesting, and I like that it will actually be used. I think that makes it one of the most interesting modules I've taken so far. It's very cool to understand the reasoning for design details with the teaching team that actually built it."}
\end{center}
\begin{center}
    \textit{``Participation in an actual to–be–deployed software project is exciting and makes your effort somewhat worthwhile."}
\end{center}

\subsection{Project Preferences}
In the first instance of our course, we allowed students to pick projects on their own. Therefore, we ended up with an imbalanced selection of projects. Students tended to prefer a project with more explicit requirements, e.g., a \parser component, instead of projects that involve more research. In the second instance, we therefore only allowed bidding on projects while the teaching team made the final decision.

\subsection{Mentoring support for third-year students}
In the second instance of our course, we had dedicated, experienced mentors (i.e., graduate students) to help student groups organize their efforts. We experienced that the additional mentorship helped SE students to quickly hands-on and get the best out of their projects. This is not only helpful to improve our system but also creates a better project experience.

\subsection{Capability of built ITS}
    
We also experienced another challenge in our journey, which was to assist students who were unable to start a task. When a student lacks a basic understanding of the problem or the necessary programming constructs, the system struggles to offer step-by-step guidance because its feedback relies on existing code structures. To address this challenge, we plan to integrate more pedagogical research with LLM into ITS to help students across different stages of learning.


\subsection{Managing Software Evolution}
Overall, we experienced that our general approach is feasible and helps both the third-year and the first-year students. However, we have also seen that we must invest significant time from our side in managing the software evolution. This includes selecting and integrating the best projects, maintaining the code base, and continuously updating the design to cater to new requirements as projects in the SE course.




\section{Impact and Vision for the future}
\label{sec:impact}

In this work, we presented our concept for linking the teaching of SE projects with the teaching of programming and introduced our Intelligent Tutoring System (ITS). Further, we discussed our experiences and observations of using the ITS through controlled experiments with first-year and second-year students and also shadow deployment with tutors. 

Based on our experience, the presented ITS impacts several aspects of programming. With our long-running teaching effort, we incrementally develop and improve the ITS into a usable product. We change how \textit{students} might learn programming and support \textit{teachers} in the introductive CS courses. Furthermore, we provide the platform for senior \textit{students} to practice software engineering in a realistic scenario. Additionally, they are encouraged to work on research-oriented topics by selecting the corresponding projects. Moreover, the ITS helps to integrate the latest \textit{research} in educational APR and related topics. Our teaching innovation can also impact students from other universities as they adopt our concept and join the ITS development team. In fact, we have already exported the teaching concept to Monash University.
 
With the shift from manual programming to AI-assisted programming, CS education must also be innovated. We think the ITS represents a well-suited platform to help students learn an effective way of using AI-based code generation tools like GitHub Copilot and ChatGPT in the future.
Therefore, instead of exposing the student directly to the AI assistant, the ITS can \textit{moderate} the prompts and explain the generated code with the support of program analysis byproduct, achieving a three-way interaction between the student, ITS, and AI assistant.

\section*{Data Availability}
Our teaching materials can be found at \url{https://cs3213-fse.github.io/lecture.html}. Our ITS is available through API requests and customized deployment via university collaboration.

\section*{Acknowledgments}
We would like to thank all students, tutors, and lecturers who have supported our work in various ways, for multi-year offerings of the course from 2021-2024 at NUS. This work was partially supported by a Singapore Ministry
of Education (MoE) Tier 3 research grant ”Automated Program Repair”, MOE-MOET32021-0001.

\bibliographystyle{IEEEtran}
\bibliography{references}

\end{document}